\documentclass{singlecol-new}

\usepackage{natbib}
\usepackage{url}
\usepackage{graphics}
\usepackage{graphicx}

\theoremstyle{TH}{
\newtheorem{theorem}{Theorem}[section]

\newtheorem{definition}[theorem]{Definition}

}

\theoremstyle{THhit}{

}

\theoremstyle{THrm}{

}

\makeatletter
\def\theequation{\arabic{equation}}
\makeatother

\begin{document}

\setcounter{page}{1}

\LRH{S. Srihari and H. W. Leong}

\RRH{Detecting sparse complexes from yeast PPI networks}

\VOL{x}

\ISSUE{x}

\BottomCatch

\PAGES{xxxx}

\CLline

\PUBYEAR{2011}

\subtitle{}

\title{Employing functional interactions for characterization and detection of sparse complexes from yeast PPI networks}

\authorA{Sriganesh Srihari\vs{-3}}

\affA{Department of Computer Science,\\
National University of Singapore,\\
Singapore 117590\\ E-mail: srigsri@comp.nus.edu.sg}

\authorB{Hon Wai Leong*\vs{-7}}

\affB{Department of Computer Science,\\
National University of Singapore,\\
Singapore 117590\\
E-mail: leonghw@comp.nus.edu.sg\\ {*}Corresponding author\vs{-2}}

\begin{abstract}
Over the last few years, several computational techniques have been devised to recover protein complexes from the protein
interaction (PPI) networks of organisms. These techniques model ``dense" subnetworks within PPI networks as complexes.
However, our comprehensive evaluations revealed that these techniques fail to reconstruct many `gold standard'
complexes that are ``sparse" in the networks
(only $71$ recovered out of $123$ known yeast complexes embedded in a network of $9704$ interactions among $1622$ proteins).
In this work, we propose a novel index called Component-Edge (CE) score to quantitatively measure the notion of ``complex
derivability" from PPI networks. Using this index, we theoretically categorize complexes as ``sparse" or ``dense"
with respect to a given network.
We then devise an algorithm SPARC that selectively employs
functional interactions to improve the CE scores of predicted complexes, and thereby elevates many of the ``sparse" complexes
to ``dense". This empowers existing methods to detect these ``sparse" complexes.
We demonstrate that our approach is effective in reconstructing significantly many complexes missed
previously (104 recovered out of the 123 known complexes or $\sim$47\% improvement).
Availability: http://www.comp.nus.edu.sg/$\sim$leonghw/MCL-CAw/
\end{abstract}

\KEYWORD{Sparse complexes; complex prediction; protein interaction networks; functional interactions.}

\maketitle

\section{Introduction}
\label{Sec_Intro}

Stoichiometrically stable complexes are formed by proteins that \emph{physically} interact to achieve
biological functions within the cell. These complexes interact with individual proteins or other complexes to form 
functional modules and pathways that drive the cellular machinery. Therefore, a faithful reconstruction of the entire set of 
complexes is essential to not only understand complex formations, but also the higher level organization of the cell.

Recent advances in high-throughput techniques have enabled to catalogue enormous amounts of physical interaction
data particularly in organisms such as \emph{Saccharomyces cerevisiae} (budding yeast).
Typically these interactions are arranged in the form of a protein interaction network (or PPI network) and 
mined for complexes using computational techniques.
From a topological perspective, these complexes are typically interpreted as regions
in the network where proteins are densely connected to each other than to the rest of the network~\citep{Zhang2008}.
Accordingly, several computational methods 
have been proposed that depend primarily on the \emph{topologies} of PPI networks,
and model \emph{dense} regions as complexes; for a survey, see~\h{\citep{Li2010,Srihari2010}}.
For example, MCL~\citep{Enright2004} simulates a series of random walks (called a \emph{flow}),
the principle being that when the walks reach a dense region, with high probability, 
they will continue to remain in that region. 
By repeated iterations of inflation (thickness) and expansion (spread) of the flow,
MCL identifies complexes. MCODE~\citep{Bader2003}, on the other hand, identifies ``seed" proteins in the
network using clustering coefficients and greedily expands in the neighborhood of these seeds 
to build complexes. CMC~\citep{Liu2009} first generates maximal
cliques from the network, and then merges highly interconnected cliques to assemble complexes. HACO~\citep{Wang2009}
performs agglomerative clustering by generating small clusters and hierarchically merging them into complexes. 
HACO improves upon the traditional hierarchical agglomerative clustering (HAC) by 
allowing for overlaps among the generated complexes.
Finally, MCL-CAw~\citep{Srihari2010} produces initial clusters using MCL and then refines these clusters by incorporating
core-attachment structure to generate complexes.

We performed comprehensive evaluations~\citep{Srihari2010} of these methods, particularly MCL, MCL-CAw, CMC and HACO,
on a variety of yeast PPI networks ranging from raw to highly-filtered and 
under varying levels of natural as well as artificial noise, and found that these methods
failed to detect many known complexes catalogued in the MIPS~\citep{Mewes2006} database.
For example, MCL missed 65 out of the 123 MIPS complexes
present in the Consolidated$_{3.19}$ network from \cite{Collins2007}. 
Even the ``union" of these methods missed 52 out of the 123 complexes.
Since the goal here is to study genome-wide compositions of complexes (the ``complexosome"), 
failure to detect even the known subset of complexes reflects severe limitations in current methods.

\subsection{Insights into the topologies of undetected complexes}
\label{Sec_Insights}
In order to understand the characteristics of these missed complexes, we ``superimposed" yeast complexes taken
from the MIPS benchmark~\citep{Mewes2006} onto the high-confidence Consolidated$_{3.19}$ yeast PPI network~\citep{Collins2007}
(\#proteins: 1622, \#interactions: 9704, average node degree: 11.187).
This ``superimposition" involves identifying the proteins of the benchmark complex in the PPI network, and extracting
out the subnetwork induced by those proteins.
Figure 1 in Supplementary materials shows this ``superimposition" visualized using \emph{Cytoscape}~\citep{Shannon2003}.

The immediate observation, which is of course typical to most PPI networks, was that the network
comprised of one main large component and multiple \emph{disjoint} smaller components of sizes 2 to 50.
Out of the 123 MIPS complexes containing at least four proteins in the network, 89 were completely
embedded in the main component, and the remaining 34 were ``scattered" among more than one components.
When we ran MCL on this network, it was able to recover only 58 of these 123 complexes.
Of the 65 undetected complexes, 27 complexes were the ones that were ``scattered",
and 34 complexes, though intact, had very low interaction densities ($<0.50$) in the network.
In fact, some of these complexes lacked internal connectivities to an extent that it was
impossible for \emph{any} algorithm to assemble back these disconnected pieces into whole complexes
solely based on topological information.
For example, the MIPS complex 510.190.110 (CCR4 complex) had seven proteins in the network scattered
among four disjoint components.
(shown within ellipses in Figure 1 in Supplementary materials). 
This complex remained disconnected with a low density of 0.1905, and naturally went undetected by all the four algorithms
(a few more examples are available from Supplementary materials).

Further, most MIPS complexes being small (sizes $\leq$ 10-15), lacking in 
just a few proteins or interactions easily rendered many complexes disconnected or with low interaction densities,
resulting in them going undetected.
All these findings revealed that a potentially strong correlation existed between the ``network constitution" of a complex
(the number of member proteins in the network and their connectivities)
and the possibility of it being detected using existing algorithms.

This work is strongly motivated by the limitations in existing complex detection methods in 
successfully detecting complexes, and the aforesaid revelations on the topologies of
these undetected complexes within PPI networks.
The purpose of our work therefore is two-fold: 
(i) to \emph{characterize} these undetected complexes, that is, to quantitatively measure their ``network constitution"; and
(ii) to propose a novel algorithm employing functional interactions to enhance the ``derivability" of sparse complexes,
which in turn empowers existing methods in detecting these complexes satisfactorily.


\section{Methods}
We represent our PPI network as $G=(V,E)$, where $V$ is the set of proteins and $E$ is the set of interactions
between the proteins. Each interaction $e = (u,v) \in E$ is assigned a weight $0 \leq w(u,v) \leq 1$ 
that reflects the confidence of the interaction, which is usually determined using an affinity weighting scheme
(the weight it is set to 1 if no scheme is used).
For any $u \in V$, $\mathcal{N}(u)$ refers to the set of neighbors of $u$.
Let $\mathcal{B} = \{B_1, B_2, ..., B_m \}$ be the set of benchmark complexes.

We propose the term \emph{sparse complexes} for the undetected complexes and ``very broadly" define them as follows:
\begin{definition}
\label{sp1}
{\sc Sparse complexes:} Given a PPI network $G$ and a set of benchmark complexes $\mathcal{B}$ known to be embedded in $G$, 
the subset $\mathcal{B}' \subseteq \mathcal{B}$ of complexes that cannot be satisfactorily detected from $G$ by existing methods are called sparse complexes.
\end{definition}

\subsection{Indices for complex derivability from PPI networks}
We next propose \emph{indices} that measure the ``derivability" of a benchmark complex from a given PPI network.
These indices capture whether or not a benchmark complex is derivable from a given PPI network, and if so, to what extent.
We propose two kinds of indices here. The first kind defines definitive criteria to categorize a given benchmark complex as 
derivable or not from the PPI network, and provides \emph{derivability bounds} on the number of such complexes in the network.
The second kind does not strictly categorize the benchmark complex as derivable or not, 
but instead assigns a \emph{derivability score} to the complex.

\subsubsection{Derivability indices with bounds}
To begin with, a naive yet natural way to categorize a benchmark complex as 
\emph{derivable} from a PPI network is if it satisfies two criteria:
(i) it has sufficient number of proteins in the network; and
(ii) it is connected within the network.

We consider a benchmark complex $B_i \in \mathcal{B}$ to be \emph{$k$-protein-derivable} from $G$ 
if at least $k > 0$ of its member proteins are
present in $G$. We consider a $k$-protein-derivable complex to be \emph{$k$-network-derivable} from $G$
if these member proteins form a connected subnetwork within $G$.

\begin{definition}
{\sc $k$-protein-derivable complex:} A benchmark complex $B_i \in \mathcal{B}$ is $k$-protein-derivable from network $G=(V,E)$
if $|B_i \cap V| \geq k$, for some $k > 0$.
\end{definition}
The set of $k$-protein-derivable complexes in $G$ is represented by $D_P(\mathcal{B},G,k)$, and
the \emph{$k$-protein-derivability index} of $G$ is $|D_P(\mathcal{B},G,k)|$.

\begin{definition}
{\sc $k$-network-derivable complex:} A benchmark complex $B_i \in \mathcal{B}$ is $k$-network-derivable from $G=(V,E)$
if $|B_i \cap V| \geq k$ for some $k > 0$, and $B_i \cap V$ forms a connected subnetwork in $G$.
\end{definition}
The set of $k$-network-derivable complexes in $G$ is represented by $D_N(\mathcal{B},G,k)$, and
the \emph{$k$-network-derivability index} of $G$ is $|D_N(\mathcal{B},G,k)|$.

\subsubsection{Derivability indices with scores}

From our extensive experiments (details omitted due to lack of space), we found that two factors strongly
contributed to the ``derivability" of a given complex from the network - the presence of a significant fraction of
complex proteins within the same connected component, and the density of the complex relative to its local neighborhood.
Based on these two factors we next define indices that assign \emph{derivability scores} to each benchmark complex to reflect the 
confidence or extent to which the complex is derivable from the network.

\emph{Component Score} $CS(B_i,G)$:
In the network $G$, let any $k$-protein-derivable complex $B_i$ be decomposed into several connected components,
$\{S_1(B_i,G), S_2(B_i,G),...,S_r(B_i,G)\}$, ordered in non-increasing order of size.
We define $CS(B_i,G)$ as the fraction of proteins within the maximal component $S_1(B_i,G)$ among 
all \emph{non-isolated proteins} in $B_i$:
\begin{equation}
CS(B_i,G) = \frac{|S_1(B_i,G)|}{|B_i'|} \textrm{ for } |B_i'| > 0, \textrm{ else } CS(B_i,G) = 0,
\end{equation}
where $B_i' = \{p: p \in B_i, \exists q \in B_i, (p,q) \in E\}$.

\emph{Edge Score} $ES(B_i,G)$:
We define $ES(B_i,G)$ as the ratio of the weight of interactions within $B_i$ to
the total weight of interactions within $B_i$ and its immediate neighborhood in $G$:
\begin{equation}
ES(B_i,G) = \frac{  \sum_{e \in E(B_i)}{w(e)}  }{  \sum_{e \in E(NB_i)}{w(e)}  } \textrm{ for } E(NB_i) \neq \emptyset, \textrm{ else } ES(B_i,G) = 0.
\end{equation}
The denominator is the weight of interactions in
the subnetwork of $G$ induced by the member proteins of $B_i$ and their direct neighbors, given by:
$V(NB_i) = \{p: p \in B_i\} \bigcup \{q: q \in \mathcal{N}(p), p \in B_i\}$ and
$E(NB_i) = \{(p,q): p,q \in V(NB_i), (p,q)\in E \}$.
Note that the edge score is different from the absolute \emph{edge density} of $B_i$, which is defined as:
$d(B_i,G) = \sum_{e \in E(B_i)}{w(e)}/(|V(B_i)|.(|V(B_i)|-1))$.

We define the \emph{Component-Edge score} $CE(B_i,G)$ as the product of the component and edge scores of $B_i$:
\begin{equation}
\label{CE}
CE(B_i,G) = CS(B_i,G)*ES(B_i,G).
\end{equation}

\begin{definition}
{\sc $k$-ce-derivable complex:} 
Given a threshold $0 \leq t_{ce} \leq 1$, a $k$-protein-derivable complex $B_i$ is $k$-$CE$-derivable if
$CE(B_i,G) \geq t_{ce}$.
\end{definition}
Therefore, the set of $k$-$CE$-derivable complexes in $G$ is given by:
$D_{CE}(\mathcal{B},G,k,t_{ce}) = \{B_i: B_i \in D_P(\mathcal{B},G,k), CE(B_i,G) \geq t_{ce}\}$,
and the \emph{$k$-CE-derivability index} of $G$ is $|D_{CE}(\mathcal{B},G,k,t_{ce})|$.

\subsubsection{Relationships among the derivability indices}
For any $k > 0$, by definition $D_N(\mathcal{B},G,k) \subseteq D_P(\mathcal{B},G,k)$.
Given a threshold $0 \leq t_{ce} \leq 1$, the relationships between $D_P(\mathcal{B},G,k)$ and $D_N(\mathcal{B},G,k)$
with $D_{CE}(\mathcal{B},G,k,t_{ce})$ are as follows. 
When $t_{ce} = 0$, all $k$-$CE$-derivable complexes are also $k$-protein-derivable, but because they may not be connected
we can say,
$D_N(\mathcal{B},G,k) \subseteq D_{CE}(\mathcal{B},G,k,t_{ce}=0) \subseteq D_P(\mathcal{B},G,k)$.
When $t_{ce} = 1$, all $k$-$CE$-derivable complexes are connected complexes that are disjoint,
therefore $D_{CE}(\mathcal{B},G,k,t_{ce}=1) \subseteq D_N(\mathcal{B},G,k) \subseteq D_P(\mathcal{B},G,k)$.
Intuitively, $t_{ce}$ can be varied in the entire range $[0,1]$ to include the ``hardest" complexes to detect
(without any internal connectivities) to only the ``easiest" complexes to detect (disjoint connected complexes). These
``hardest" complexes to detect can form ``holes" in the network by having zero interactions among their member proteins
but having interactions with their immediate neighbors
(see Supplementary materials for a visual representation of these complex sets).

\subsubsection{Validating the derivability indices against ground truth}
We now validate the derivability scores ($CS$, $ES$, $CE$ scores and absolute edge density)
of benchmark complexes with respect to the PPI network against the accuracies with which these complexes are
actually derived using existing methods.
This will reveal how effective each of these indices are in capturing actual complex derivability
using existing methods.

\begin{table}[ht]
\caption{Comparing $CE$-score with edge density:
Correlation between the edge density / $CE$-scores of MIPS complexes and their Jaccard accuracies
when actually derived from the Consolidated network using MCL.}
\begin{center}
{\small
\begin{tabular}{ l || c || c | c | c }
\multicolumn{5}{c}{\emph{The Consolidated$_{3.19}$ network: \#p 1622, \#i 9704}}\\[0.8ex]
\hline
			&	\multicolumn{4}{c}{Pearson correlation}								\\
			&	\multicolumn{4}{c}{with Jaccard accuracy}							\\[0.8ex]
\hline
			&					&	\multicolumn{3}{c}{Our indices}							\\[0.8ex]
{Method}		&	{Edge density}			&	{$CE$}			& 		{$CS$}		& {$ES$}		\\[0.8ex]
\hline
MCL			&	\emph{0.101}			&	\emph{0.719}		& 		0.511		& 0.518			\\
MCL-CAw			&	\emph{0.196}			&	\emph{0.785}		& 		0.492		& 0.628			\\
CMC			&	\emph{0.174}			&	\emph{0.649}		& 		0.471		& 0.477			\\
HACO			&	\emph{0.159}			&	\emph{0.786}		& 		0.472		& 0.608			\\

\hline	
\end{tabular}
}
\end{center}
\label{table:Derivability_indices_correlation_Consol}
\end{table}

We use two PPI networks for this validation, the Consolidated$_{3.19}$ network (a weighted network)
from~\cite{Collins2007}, and 
the `Filtered Yeast Interaction' (FYI) network (a literature-validated but unweighted network) from~\cite{Han2004}.
We use complexes from the MIPS and Wodak catalogues as our benchmark complexes.
Table~\ref{table:Derivability_indices_correlation_Consol} shows the Pearson correlation values between the derivability scores
and the \emph{Jaccard} accuracies obtained from four complex detection methods, MCL, MCL-CAw, CMC and HACO
(the complete set of results are available from the Supplementary materials).
The results show the $CE$-scores and Jaccard accuracies are \emph{strongly correlated} (Pearson: 0.719 using MCL), better
than the correlation between absolute edge densities and Jaccard accuracies (Pearson: 101 using MCL).
This means our proposed $CE$-score is a \emph{stronger} indicator of actual complex derivability compared to
the traditionally adopted indicators like edge density.
(Even the individual scores, $CS$ and $ES$ show reasonable correlation with Jaccard accuracies.
Also, there are a few other indices like global and local modularity~\citep{Newman2004}, but
these do not capture the notion of proteins being part of the same connected component, 
and they perform similar to our edge-score $ES$).

\subsection{A measure of sparse complexes}
We can now employ our proposed $CE$-score to give a more quantitative definition for sparse complexes.
\begin{definition}
\label{sp2}
{\sc Sparse complexes:} Given a PPI network $G$, a benchmark complex $B_i$ and a threshold $0 \leq t_{ce} \leq 1$, 
the complex $B_i$ is called sparse with respect to $G$ if $CE(B_i,G) < t_{ce}$.
\end{definition}

Notice how the two definitions~\ref{sp1} and~\ref{sp2} can be ``linked" using our $CE$-score and threshold $t_{ce}$,
which offer a quantitative value to the derivability of complexes.
If this value is less than a certain threshold, the complex
is highly likely to go undetected from existing methods and therefore it is \emph{sparse}, 
else it is highly likely to be detected and therefore it is \emph{dense}.
In general, for the benchmark complexes $\mathcal{B}$,
the set of sparse complexes is given by 
$\mathcal{S}(\mathcal{B}, G, k, t_{ce}) = \{B_i: B_i \in D_P(\mathcal{B},G,k), CE(B_i,G) < t_{ce}\}$, 
and its complementary set
$\mathcal{D}(\mathcal{B}, G, k, t_{ce}) = \{B_i: B_i \in D_P(\mathcal{B},G,k), CE(B_i,G) \geq t_{ce}\}$
forms the dense complexes.
The threshold $t_{ce}$ defines this ``boundary" between the sparse and dense benchmark complexes in the network.
Since we do not know at which value of $t_{ce}$ existing methods operate, we propose an approach that ``packs"
higher number of dense complexes for all values of $t_{ce} \in [0,1]$ or at least for the larger values of $t_{ce}$.


\subsection{Detecting sparse complexes}
We noted in Section~\ref{Sec_Insights} that existing methods are severely constrained by 
``gaps" in crucial topological information required to ensure the
two required criteria for complex derivability namely, component-based connectivity and relative edge density.
In fact, any new method based solely on PPI networks would also face these constraints.
Due to these reasons, a natural approach to aid existing methods or devise new methods 
would be to first fill these ``topological gaps" in existing PPI networks.

Even though this seems like a simple enough solution to pursue, 
we are severely lacking in the interaction data required to fill these gaps.
Current estimates on yeast~\citep{Cusick2008}, put the verified fraction of the physical interactome to $\sim$70\%,
which means we are still lacking in $\sim$30\% reliable interaction data, mainly due to limitations
in existing experimental and computational techniques.
Consequently, a novel solution is to look beyond physical interactions to fill these topological
gaps. In our work, we propose to use \emph{functional interactions} for this purpose, specifically
aimed at improving complex prediction.

\subsubsection{Employing functional interactions to detect sparse complexes}
Functional interactions or associations are logical interactions among proteins that 
share similar functions~\citep{String2003}. 
These interactions can be inferred among proteins participating in the
same multi-protein assemblies (complexes, functional modules and pathways),
or annotated to similar biological functions and processes, 
or encoded by genes maintained and regulated together or genes having the same `phylogenetic profile'
(present or absent together across several genomes), etc.~\citep{String2003}.
Therefore, these interactions ``encode" information beyond just direct physical interactions.
In fact many of the computational methods developed to predict protein interactions mainly manage to predict
functional interactions.

Functional interactions can be considered more ``general" or a ``superset" of direct physical interactions:
two proteins involved in a stable physical interaction are functionally related, but two proteins involved in a functional
interaction may not necessarily interact physically.
This means functional interactions have a potential to effectively \emph{complement} physical interactions.
We capitalize on this complementarity by non-randomly adding functional interactions to ensure the two required criteria:
(i) Some functional interactions may be direct physical interactions missing in the physical datasets - 
these are directly useful to ``pull-in" disconnected proteins; and
(ii) Even if some functional interactions do not correspond to direct physical interactions, if they fall
within the same complex, they can ``artificially" increase the density of that complex.

\subsubsection{The SPARC algorithm for employing functional interactions}
Here, we propose a post-processing based algorithm SPARC
to empower existing methods in detecting SPARse Complexes by using functional interactions. SPARC works as follows.
Let $G_P = (V_P, E_P)$ be the PPI network and $G_F = (V_F, E_F)$ be the functional network.

\emph{Step 1:} The input to the algorithm is the set of physical clusters $\mathcal{C}_P$
from network $G_P$ generated using an existing method. It then calculates the $CE$-score
$CE(G_P,C_i)$ for each cluster $C_i \in \mathcal{C}_P$.
All clusters with $CE$-scores above a threshold $\delta$, that is,
$\{C_i \in \mathcal{C}_P: CE(C_i, G_P) \geq \delta \}$, are output as predicted complexes, 
while the remaining are reserved for further processing.

\emph{Step 2:} We then add-in the interactions of $G_F$ to $G_P$ to produce
a larger network $G_A = (V_A, E_A)$, where $V_A = V_P \cup V_F$ and $E_A = E_P \cup E_F$.

\emph{Step 3 (iterative):} 
For each reserved cluster $C_j$, the $CE$-score is recalculated with respect to $G_A$.
If for the cluster $C_j$, the $CE$-score improves beyond $\delta$, that is, $CE(C_j,G_A) \geq \delta$,
it is output as a predicted complex.
If not, we explore in the neighborhood of $C_j$ to include proteins that can potentially improve
$CE(C_j,G_A)$. We consider the set of direct neighbors $\mathcal{N}(C_j,G_A)$, and
sort them in non-increasing order of their interaction weights to $C_j$.
We then repeatedly consider a protein $p \in \mathcal{N}(C_j,G_A)$ in that order
such that $CE(C_j \cup \{p\}, G_A) > CE(C_j,G_A)$ and add it to $C_j$, till the $CE$-score cannot be improved any further.
If the improved $CE$-score manages to cross $\delta$, we output the cluster $C_j$ as a predicted complex.

The key idea behind SPARC is as follows. Many complexes have low $CE$-scores in the PPI network. If adding
functional interactions can either increase their internal connectivities or ``pull in" 
the disconnected proteins, we can increase the $CE$-scores of these complexes. However, blindly adding functional interactions
can result in many false positive predictions. Therefore, here we selectively utilize functional interactions only to improve the
$CE$-scores of clusters predicted out of the physical network. Those clusters that show the improvement correspond to
real complexes.


\section{Experimental results and discussion}

\subsection{Preparation of experimental data}
We gathered physical interactions from \emph{Saccharomyces cerevisiae} (budding yeast) inferred from the 
following yeast two-hybrid and affinity purification experiments, deposited in Biogrid~\citep{Biogrid2003}:
Uetz (2000), Ito (2001), Gavin (2002, 2006), Krogan (2006), Collins (2007) and Yu (2008),
to build the protein interaction network, which we call the \emph{Physical network} $P$. The interactions of $P$
are not scored.

Next, high-confidence functional interactions from yeast were gathered from the String database~\citep{String2003} 
to build the \emph{Functional network} $F$. These functional interactions showed confidence scores $\geq$ 0.90 
in at least two of the following evidences: gene neighborhood, co-occurrence, co-expression and text mining
(these scores are available from String).

We combined the two networks to generate a larger network which we call the 
\emph{Augmented} Physical+Functional network $P+F$.
Table~\ref{table:PPI_Network_Properties} shows some properties of these networks.
The overlaps between $P$ and $F$ are as follows: $|V(P) \cap V(F)| = 2928$ and $|E(P) \cap E(F)| = 1296$.

\begin{table}[ht]
\caption{Properties of the physical and functional networks obtained from yeast.}
  \begin{center}
  {\small
  \begin{tabular}{ l | c | c | c }
  \hline
		{Network}	 			& {\# Proteins} & {\# Interactions} &   {Avg node degree} \\[0.8ex]
  \hline
	
		Physical $(P)$				& 4113		 & 26518		& 12.89	\\[1.0ex]
  
		Functional $(F)$			& 3960		 & 18683		& 10.12	\\[1.0ex]	
  
  		Augmented $(P+F)$			& 5145		& 43905			& 17.07	\\[1.0ex]  			
  \hline
  \end{tabular}
  }
  \end{center}
  \label{table:PPI_Network_Properties}
  \end{table}

The presence of \emph{noise} (false positives) is a severe limiting factor in publicly available interaction datasets
in spite of gathering only high-confidence datasets. Therefore, we further \emph{filtered} these datasets, which
involves assigning each interaction a confidence score (between 0 and 1) that reflects its reliability,
and discarding interactions with low scores ($< 0.20$).
Here, we (re)scored the networks using three scoring schemes, two of which were based on network topology namely,
\emph{FS-Weight} devised by~\cite{Chua2008} and \emph{Iterative-CD} devised by~\cite{Liu2009},
while the third was based on evidences from Gene Ontology (GO)~\citep{Ashburner2000},
called \emph{TCSS} devised by~\cite{Jain2010}.


\subsubsection{Benchmark complexes and GO annotations}
The \emph{benchmark} or reference set of complexes was assembled from two sources:
313 complexes of MIPS~\citep{Mewes2006} and
408 complexes of the Wodak lab CYC2008 catalogue~\citep{Pu2009}.
The properties of these benchmark sets are shown in Table~\ref{table:Size_distribution_benchmarks}.
For the evaluation, we considered only the 4-protein-derivable complexes out of these sets.
This is because it is typically difficult to predict very small complexes (size $< 4$) with high accuracy
by using primarily topological information~\h{\citep{Liu2009,Srihari2010}}.

 \begin{table}[ht]
 \caption{Properties of hand-curated (benchmark) yeast complexes from the MIPS and Wodak CYC2008 catalogues.}
  \begin{center}
  {\small
  \begin{tabular}{ c | c | c  c  c  c }
  \hline
  			&			& \multicolumn{4}{c}{Size distribution} 		\\
  {Benchmark}		& {\#Complexes} 	& $<$ 3  & 3-10		& 11-25		& $>$ 25	\\
  \hline
  MIPS			& 313			& 106    & 138		& 42		& 27		\\
  Wodak			& 408			& 172 	 & 204		& 27		& 5		\\  	    
  \hline

\end{tabular}
}
\end{center}
\label{table:Size_distribution_benchmarks}
\end{table}

The GO annotations for yeast proteins were downloaded from 
the \emph{Saccharomyces} Genome Database (SGD)~\citep{Cherry98}, 
which include the annotations (not considering the Inferred from Electronic Annotations or IEA)
for three ontologies - Cellular Component (CC), Biological Process (BP) and Molecular Function (MF).
These annotations were used as evidences in the TCSS scheme~\citep{Jain2010}.
We excluded the branch corresponding to the GO term `macromolecular complex' (GO:0032991) 
to avoid any bias coming from the GO complexes.

\subsection{Complex detection algorithms and evaluation metrics}
We used four complex detecting algorithms mentioned previously, MCL~\citep{Enright2004}, CMC~\citep{Liu2009},
HACO~\citep{Wang2009} and MCL-CAw~\citep{Srihari2010}.
Some of their properties and the preset parameter values are summarized in Table~\ref{Table:Evaluation_algorithms}.
These methods are different from one another in the algorithmic techniques employed, and therefore form a good mix of methods
for our evaluation.

\begin{table}[ht]
\caption{Existing complex detection methods used in the evaluation.}
\begin{center}
{\small
\begin{tabular}{ c || c | c | c | c }
\hline
{Property} 		& {\bf MCL} 			& {\bf MCL-CAw}				& {\bf CMC}			&{\bf HACO}\\[0.8ex]
\hline
{\bf Principle} 	& Flow 				& Core-attach 				& Maximal			& Hier agglo\\
                	& simulation			& refinement				& clique			& cluster with\\
                	& 				& over MCL				& merging        		& overlaps\\[1.3ex]        
                	
{\bf Parameters}        &  $I$                 		& $I$, $\alpha$, $\gamma$		& Merge $m$,			& UPGMA 		\\
{\bf (preset values)}  	& (2.5)				& (2.5, 1.5, 0.75 )			& Overlap $t$, 			& cutoff		\\
			&				&					& Min clust size		& (0.2)			\\
			&				&					& (0.5, 0.4, 4)			&			\\
\hline	
\end{tabular}
}
\end{center}
\label{Table:Evaluation_algorithms}
\end{table}

Usually, recall $Rc$ (coverage) and precision $Pr$ (sensitivity) are used to evaluate the performance of methods
against benchmark complexes. Here, we use previously reported~\citep{Liu2009} definitions for these measures.
Let $\mathcal{B} = \{B_1, B_2,...,B_m \}$ and $\mathcal{C} = \{C_1, C_2, ...,C_n \}$ be the sets of 
benchmark and predicted complexes, respectively. We use the Jaccard coefficient $J$ to quantify the overlap between 
a $B_i$ and a $C_j$: $J(B_i,C_j) = |B_i \cap C_j|/|B_i \cup C_j|$.

We consider $B_i$ to be covered by $C_j$, if $J(B_i,C_j) \geq$ {\em overlap threshold} $J_{min}$. 
In our experiments, we set the threshold $J_{min} = 0.50$, which requires $|B_i \cap C_j| \geq \frac{|B_i| + |C_j|}{3}$. 
For example, if $|B_i| = |C_j| = 8$, then the overlap between $B_i$ and $C_j$ should be at least 6.
Based on this the recall $Rc$ is given by:
\begin{equation}
\label{eq_recall}
	Rc\mathcal{(B,P)} = \frac{|\{B_i|B_i \in \mathcal{B} \wedge \exists C_j \in \mathcal{C}; J(B_i,C_j) \geq J_{min}\}|}{|\mathcal{B}|}.
\end{equation}
Here, $|\{B_i|B_i \in \mathcal{B} \wedge \exists C_j \in \mathcal{C}; J(B_i,C_j) \geq J_{min}\}|$ gives the number of {\em derived benchmarks}.
And the precision $Pr$ is given by:
\begin{equation}
\label{eq_precision}
	Pr\mathcal{(B,P)} = \frac{|\{ C_j| C_j \in \mathcal{C} \wedge \exists B_i \in \mathcal{B}; J(B_i,C_j) \geq J_{min} \} |}{|\mathcal{C}|}.
\end{equation}
Here, $|\{C_j| C_j \in \mathcal{C} \wedge \exists B_i \in \mathcal{B}; J(B_i,C_j) \geq J_{min}\}|$ gives the number of {\em matched predictions}.


\subsection{Impact of adding functional interactions on complex derivability}

To begin with, we measured the number of derivable benchmark complexes from the
Physical $(P)$, Functional $(F)$, Augmented $(P+F)$ networks and their scored versions,
$ICD(P+F)$, $FSW(P+F)$ and $TCSS(P+F)$, using our proposed derivability indices. 

Table~\ref{table:Derivability_indices1} shows the number of protein-derivable and network-derivable benchmark complexes 
from these networks. The findings can be summarized as follows:
(a) The network-derivable complexes were significantly fewer than the protein-derivable complexes 
further supporting the claim (Section~\ref{Sec_Intro}) that many benchmark complexes remained disconnected within the networks.
(b) The number of protein-derivable and network-derivable complexes were higher for the $P+F$ network 
than the individual $P$ and $F$ networks. The significance of this 
increase was gauged against a random network $R$ built using the same set of proteins and the average node degree in $F$.
The $P+R$ network showed fewer network-derivable complexes compared to $P+F$.
This indicated that $F$ added more interactions to ``complexed" regions in $P$ compared to what the $R$ network added.
(c) The number of protein-derivable and network-derivable complexes in the scored networks,
$ICD(P+F)$, $FSW(P+F)$ and $TCSS(P+F)$, were fewer than the $P+F$ network.
This is not a concern because filtering usually discards interaction data leading to smaller networks.
(d) Even though protein-derivable complexes in the scored networks were fewer than the $P+F$ network,
the corresponding decrease in network-derivable complexes was relatively marginal. 
This indicated that the scoring schemes retained most interactions among complexed proteins, 
and discarded mainly the noisy ones.

\begin{table}[ht]
\caption{Impact of adding functional interactions on protein-derivability and network-derivability of MIPS complexes.}
\begin{center}
{\small
\begin{tabular}{ l || c | c }
				\multicolumn{3}{c}{MIPS: \#313; $k=4$}			\\[0.8ex]
\hline
			& 		{\#Protein-} 		& 	{\#Network-} 	\\
{Network}		&		{derivable}		&	{derivable}	\\
\hline		
Physical $P$		&			155		&	59		\\
Functional $F$		&			153		&	28		\\
P+Random		&			164		&	61		\\
P+F			&			164		&	68		\\
ICD(P+F)		&			122		&	64		\\
FSW(P+F)		&			119		&	64		\\
TCSS(P+F)		&			158		&	68		\\	

\hline
\end{tabular}
}
\end{center}
\label{table:Derivability_indices1}
\end{table}

\begin{table}[ht]
\caption{Impact of adding functional interactions on $CE$-derivability of MIPS complexes.}
\begin{center}
{\small
\begin{tabular}{ c || c | c | c | c | c | c }
\multicolumn{7}{c}{MIPS: \#313; $k=4$}\\
\hline
		& \multicolumn{6}{c}{\# Complexes with $CE$-score $\geq t_{ce}$}\\			
\hline
{Threshold $t_{ce}$}	& {P}	& {F}	& {P+F}		&{ICD(P+F)}	& {FSW(P+F)}	& {TCSS(P+F)}   \\[0.8ex]
\hline
0.00			& 155	& 153	& 164		& 152		& 119		& 162		\\
0.10			& 153	& 151	& 162		& 148		& 116		& 160		\\
0.20			& 149	& 136	& 158		& 145		& 113		& 157		\\
0.30			& 140	& 108	& 149		& 142		& 110		& 154		\\
0.40			& 129	& 81	& 135		& 137		& 108		& 148		\\
0.50			& 101	& 54	& 102		& 112		& 101		& 126		\\
0.60			& 81	& 21	& 70		& 93		& 87		& 101		\\
0.70			& 62	& 9	& 55		& 71		& 69		& 86		\\
0.80			& 39	& 0	& 34		& 44		& 42		& 59		\\
0.90			& 19	& 0	& 14		& 21		& 21		& 35		\\
1.00			& 6	& 0	& 3		& 11		& 10		& 18		\\

\hline	
\end{tabular}
}
\end{center}
\label{table:CE_derivability_MIPS}
\end{table}

Next, Table~\ref{table:CE_derivability_MIPS} shows the number of $CE$-derivable benchmark complexes 
from these networks for all threshold values $t_{ce} \in [0,1]$.
This table does a more fine-scale dissection of the improvement shown before.
For lower values of $t_{ce}$, the number of $CE$-derivable complexes was higher for $P+F$ compared to $P$.
But, for higher values of $t_{ce}$, the number was lower compared to $P$. 
Similarly, for lower values of $t_{ce}$, the number of $CE$-derivable complexes was higher for $P+F$ compared
to the three scored networks. But, for higher values of $t_{ce}$, the three scored networks showed considerably higher
$CE$-derivable complexes than both the $P$ and $P+F$ networks.
These findings indicate that noise had a sizable impact on the $CE$-scores of complexes: the improvement obtained by
adding functional interactions was completely canceled out by noise, leading to lower performance of the $P+F$ network.
But, affinity scoring (filtering) considerably alleviated this impact of noise, thereby improving
the $CE$-derivability of the networks.


\subsection*{Improvement in complex detection using SPARC}

Table~\ref{table:Methods1_MIPS} shows the performance of the four methods MCL, MCL-CAw, CMC and HACO
on the raw physical and scored physical networks 
(we do not show the results on $F$ because functional interactions are only used to improve the physical clusters,
and not for complex detection by themselves - many of the functional clusters do not correspond to physical complexes). 
It shows that scoring helped to reconstruct significantly more complexes and
with better accuracies (also noted in~\h{\citet{Srihari2010}}).

Next, Table~\ref{table:Methods2_MIPS} shows the performance after refining the physical clusters using
functional interactions by applying SPARC ($\delta = 0.40$).
It shows that post-processing using raw functional interactions $(P+F)$ led to many noisy clusters, 
resulting in lower precision and recall.
But, using filtered (scored) functional interactions helped to reconstruct significantly more complexes out of the
physical clusters.

One interesting point to note is that the compositions of predicted complexes vary based on 
the scoring scheme used (also noted in~\citet{Srihari2010}), and therefore we had to 
construct a \emph{consensus set} of complexes from the three scoring schemes for each of the methods.
To do this, we employed a three-way agreement scheme based on Jaccard overlaps.
Let $\{A, B, C\}$ be a complex triplet, each complex predicted from a different scored network by the same method. 
If at least two complex pairs from $\{(A,B), (B,C), (C,A)\}$ achieve significant Jaccard overlaps $(\geq 0.70)$, 
then the proteins of $A$, $B$ and $C$ are merged together into a single consensus complex $T$.
Only the proteins originating from at least two complexes are included in $T$. 
We noticed that this consensus operation further improves the accuracies of the predictions leading to better
reconstruction of benchmark complexes.

\begin{table}[htp]
\caption{Impact of scoring on complex detection methods (evaluation against MIPS). `Derivable' refers to 4-protein-derivable complexes.}
{\scriptsize
\begin{center}
\begin{tabular}{ l || l || c | c || c | c || c | c }
			    & 			& \multicolumn{6}{c}{Matched against MIPS complexes. Jaccard threshold $J_{min}=0.50$.}																	\\

\hline			   
		{Method}    & {Network} 	& {\#Predicted}	   	& {\#Matched}	 	& {\#Derivable}		& {\#Derived}	 	 		& {$Pr$}		& {$Rc$}	 \\
\hline	    	

			    & 	Physical P	& 	294		& 	29		&   155			& 38					& 0.098			& 0.245		\\[0.8ex]			    
		MCL	    &	FSW(P)		& 	156		& 	31		&   102 		& 40					& 0.198			& 0.333		\\[0.8ex]			 			    
			    & 	ICD(P)		& 	167		& 	32		&   109			& 40					& 0.191			& 0.293 	\\[0.8ex]		
			    & 	TCSS(P)		& 	172		& 	39		&   112			& 41					& 0.226			& 0.366		\\[0.8ex]

\hline		
			    & 	Physical P	& 	297		& 	39		& 155			& 49					& 0.131			& 0.316		\\[0.8ex]			    
		MCL	    & 	FSW(P)		& 	149		& 	38		& 102			& 51					& 0.255			& 0.392		\\[0.8ex]			    
		-CAw	    & 	ICD(P)		& 	162		& 	41		& 109			& 52					& 0.253			& 0.376		\\[0.8ex]					    
			    & 	TCSS(P)		& 	168		& 	41		& 112			& 54					& 0.244			& 0.366		\\[0.8ex]

\hline
			    & 	Physical P	& 	156		&  	41		& 155			& 56					& 0.263			& 0.361		\\[0.8ex]			    
		CMC	    & 	FSW(P)		& 	144		& 	31		& 102			& 59					& 0.215			& 0.313		\\[0.8ex]			    			    
			    & 	ICD(P)		& 	165		& 	43		& 109			& 60					& 0.260			& 0.394		\\[0.8ex]		
			    & 	TCSS(P)		& 	128		& 	39		& 112			& 59					& 0.304			& 0.357		\\[0.8ex]

\hline		
			    & 	Physical P	& 	414		& 	34		& 155			& 41					& 0.082			& 0.264		\\[0.8ex]			    
		HACO	    & 	FSW(P)		& 	221		& 	32		& 102			& 44					& 0.144			& 0.313 	\\[0.8ex]			    			    
			    & 	ICD(P)		& 	248		& 	37		& 109			& 45					& 0.149			& 0.339		\\[0.8ex]		
			    & 	TCSS(P)		& 	253		&	46		& 112			& 45					& 0.181			& 0.410		\\[0.8ex]

\hline		

\end{tabular}
\end{center}
}
\label{table:Methods1_MIPS}
\end{table}

\begin{table}[htp]
\caption{Impact of adding functional interactions using SPARC on complex detection methods (evaluation against MIPS). `Derivable' refers to 4-protein-derivable complexes.}
{\scriptsize
\begin{center}
\begin{tabular}{ l || l || c | c | c || c | c || c | c }
			    & 			& \multicolumn{7}{c}{Matched against MIPS complexes. Jaccard threshold $J_{min}=0.50$.}																	\\

\hline			   
		{Method}    & {Network} 	& {\#Predicted}	   	&{Size}	 	& {\#Matched}	 	& {\#Derivable}		& {\#Derived}	 	 		& {$Pr$}		& {$Rc$}	 \\
\hline	    	

			    &   P		&       294		& 7.96		& 	29		&   155			& 38					&	0.098		& 0.245		\\[0.8ex]
			    &   P+F		& 	338		& 8.66		& 	19		&   164			& 23					& 	0.056		& 0.140		\\[0.8ex]			    
		MCL	    &	FSW(P+F)	& 	102		& 15.88		& 	29		&   119			& 38 					& 	0.284		& 0.319		\\[0.8ex]			 			    
			    & 	ICD(P+F)	& 	138		& 17.14		& 	33		&   122			& 44					& 	0.239		& 0.361		\\[0.8ex]		
			    & 	TCSS(P+F)	& 	261		& 10.52		& 	42		&   158			& 54					& 	0.161		& 0.342		\\[0.8ex]			    
			    &   Consensus	&	429		& 13.01		&	57		&   164			& 56					&	0.133		& 0.341		\\[0.8ex]

\hline		
			    &	P		&	297		& 7.94		&	39		& 155			& 49					&	0.131		& 0.316		\\[0.8ex]
			    &   P+F		& 	342		& 8.34		& 	25		& 164			& 29					& 	0.073		& 0.177		\\[0.8ex]
		MCL	    & 	FSW(P+F)	& 	136		& 9.46		& 	41		& 119 			& 57					& 	0.301		& 0.479		\\[0.8ex]			    
		-CAw	    & 	ICD(P+F)	& 	141		& 7.44		& 	48		& 122			& 61					& 	0.340		& 0.500		\\[0.8ex]					    
			    & 	TCSS(P+F)	& 	296		& 9.98		& 	49		& 158			& 61					& 	0.166		& 0.386		\\[0.8ex]			    
			    &   Consensus	&	484		& 8.72		&	81		& 164			& 71					&	0.167		& 0.432		\\[0.8ex]

\hline
			    &	P		&	156		&   11.42	&	41		&  155			& 56					&	0.263		& 0.361		\\[0.8ex]
			    &   P+F		& 	306		&   14.39	& 	33		&  164			& 41					& 	0.108		& 0.250		\\[0.8ex]			    			    
		CMC	    & 	FSW(P+F)	& 	136		&   12.44	& 	36		&  119			& 48					& 	0.265		& 0.403		\\[0.8ex]			    			    
			    & 	ICD(P+F)	& 	252		&   8.91	& 	51		&  122			& 63					& 	0.202		& 0.516		\\[0.8ex]		
			    & 	TCSS(P+F)	& 	127		&   11.66	& 	45		&  158			& 60					& 	0.354		& 0.380		\\[0.8ex]			    
			    &   Consensus	&	429		&   9.80	&	80		&  164 			& 66					&	0.186		& 0.402		\\[0.8ex]

\hline		
			    &	P		& 	414		&   5.98	&	34		& 155			&  41					&	0.082		& 0.264		\\[0.8ex]
			    & 	P+F		& 	510		&   6.68	& 	28		& 164			&  34					& 	0.055		& 0.207		\\[0.8ex]			    			    
		HACO	    & 	FSW(P+F)	& 	111		&   10.17	& 	39		& 119			&  54					& 	0.351		& 0.454		\\[0.8ex]			    			    
			    & 	ICD(P+F)	& 	131		&   8.90	& 	43		& 122			&  60					& 	0.328		& 0.492		\\[0.8ex]		
			    & 	TCSS(P+F)	& 	269		&   7.49	&	55		& 158			&  67					& 	0.204		& 0.424		\\[0.8ex]			    
			    &   Consensus	&	419		&   7.61	&	79		& 164			&  74					&	0.189		& 0.451		\\[0.8ex]

\hline		

\end{tabular}
\end{center}
}
\label{table:Methods2_MIPS}
\end{table}


Finally, Table~\ref{table:Methods3_MIPS} compares the number of benchmark complexes successfully reconstructed by sparse clusters
before and after the SPARC-based post-processing. It clearly demonstrates that many physical clusters were in fact
sparse ($CE$-score $< 0.40$), many of which underwent post-processing by SPARC. These post-processed clusters
were able to reconstruct significantly higher number of benchmark complexes.
Figure 5 in Supplementary materials correlates the improvement in $CE$-scores of these sparse clusters
with the improvement in their Jaccard accuracies when matched to benchmark complexes.

\begin{table}[htp]
\caption{
The number of benchmark complexes recovered by sparse clusters before and after the SPARC-based
processing.}
{\scriptsize
\begin{center}
\begin{tabular}{l || l || c || c | c || c || c | c }
\hline
			& 		   &\multicolumn{4}{c ||}{\#Predicted clusters}		&									\multicolumn{2}{c}{\#Benchmarks}	\\[0.8ex]
			&		   &				& {Sparse}		& 			&{Final}		& {Derived}		& {Derived}		\\
{Method}		&	{Network}  &	{Initial}		& {($CE < 0.40$)}	& {Processed}		&{(Size $\geq 4$)}	& {(Before)}		& {(After)}		\\[0.8ex]
\hline			
			&	$P$	   &	638			&  269			&  8			& 338			& 0			& 2			\\[0.8ex]
	
MCL			&       FSW(P+F)   &	188			&   42			&  16			& 102			& 1			& 9			\\
			&	ICD(P+F)   &    258			&   57			&  18			& 138			& 2			& 9			\\
			&	TCSS(P+F)  &    380			&   102			&  19			& 261			& 2			& 10			\\[0.8ex]
\hline
			&	$P$	   &	472			&  212			&  8			& 342			& 0			& 2			\\[0.8ex]
	
MCL-			&       FSW(P+F)   &	255			&   37			&  19			& 136			& 2			& 11			\\
CAw			&	ICD(P+F)   &    258			&   39			&  21			& 141			& 2			& 13			\\
			&	TCSS(P+F)  &    408			&   97			&  26			& 296			& 3			& 16			\\[0.8ex]
\hline
			&	$P$	   &	424			&  186			&  20			& 306			& 0			& 8			\\[0.8ex]
	
CMC			&       FSW(P+F)   &	251			&   32			&  23			& 136			& 2			& 18			\\
			&	ICD(P+F)   &    354			&   44			&  36			& 252			& 2			& 21			\\
			&	TCSS(P+F)  &    224			&   56			&  41			& 127			& 4			& 27			\\[0.8ex]
\hline
			&	$P$	   &	389			&  25			&  510			& 338			& 1			& 10			\\[0.8ex]
	
HACO			&       FSW(P+F)   &	53			&   29			&  111			& 102			& 2			& 21			\\
			&	ICD(P+F)   &    59			&   31			&  131			& 138			& 3			& 23			\\
			&	TCSS(P+F)  &    66			&   43			&  269			& 261			& 6			& 36			\\[0.8ex]
\hline

\end{tabular}
\end{center}
}
\label{table:Methods3_MIPS}
\end{table}


\subsection*{Some case studies of detected complexes}
We performed in-depth analysis of some of the predicted complexes using
\emph{Cytoscape}~\citep{Shannon2003}.
For example, the CCR4-NOT complex is a multifunctional
complex that regulates transcription, plays a role in mRNA degradation, and also regulates
cellular functions in response to changes in environmental signals in yeast~\citep{Panasenko2006}.
This complex was ``scattered" among multiple disjoint components of the Physical network, and therefore
went undetected from all four methods. The addition of functional interactions
facilitated linking together of these components,
enabling the methods to detect it successfully (see Figure 1).

\begin{figure}
\begin{center}
\includegraphics[scale=0.75]{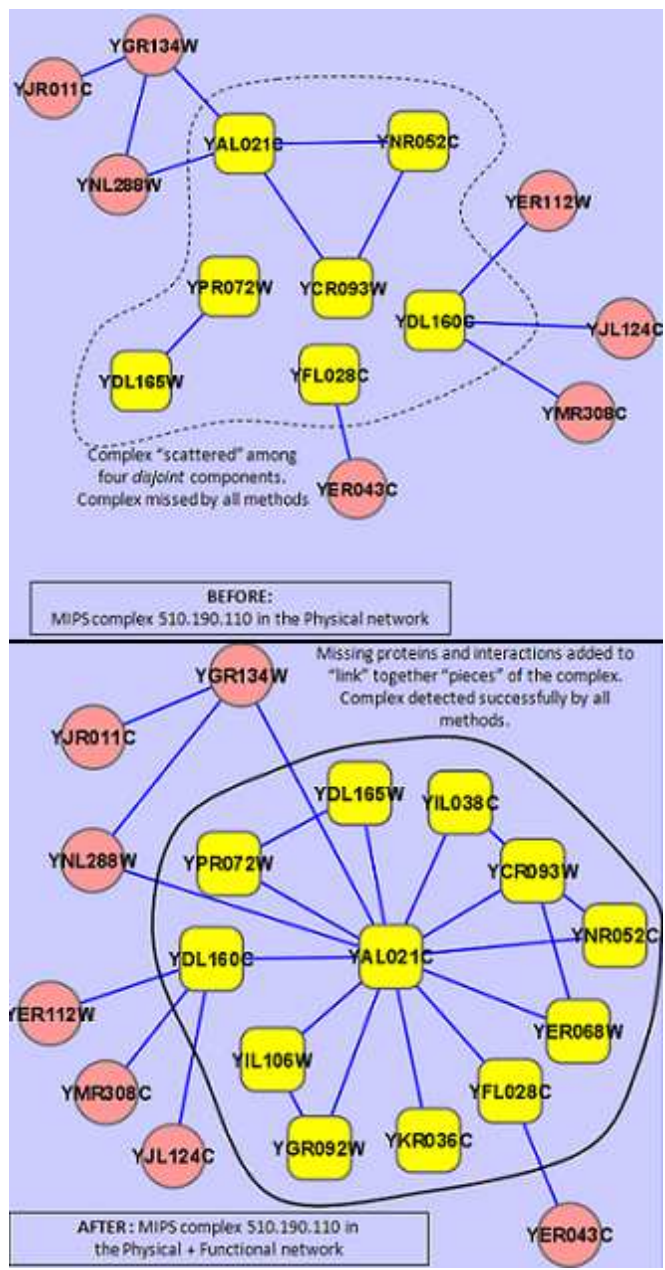}
\label{figure:510.190.110.before_after_shrink}
\end{center}
\caption{MIPS 510.190.110 complex before and after refinement using functional interactions by SPARC, and the effect
on its detection using existing methods. BEFORE: The complex was ``scattered" among four components; $CE$-score = 0.1905. 
AFTER: The four components were linked together into a single component; $CE$-score = 0.623.}
\end{figure}

While many additional complexes were detected using SPARC-based refinement, 
there were a few complexes that were missed as well (see Supplementary materials for a list).
For example, the RNA polymerase complexes I, II and III, that are involved in the formation of RNA chains
during transcription~\citep{Hurwitz2005}, were bundled into a large dense module together with some of the
TBP-associated factors and TFIID complexes, which are also involved in transcription~\citep{Green2000}. 
Due to the functional similarity between the subunits of all these complexes,
several functional interactions were added among them. Consequently, the methods recovered a large
dense module housing all these complexes from which the individual complexes could not be segregated.
The same was the case with the multi-eIF complexes and the SAGA-SLIK-ADA-TFIID complexes.
The increase in the average cluster sizes in Table~\ref{table:Methods2_MIPS} further depict this effect.

\subsection*{Discussion}
Functional interactions can be considered a ``superset" of physical interactions.
However, the low overlaps between the $P$ and $F$ networks seems to be projecting a suprisingly different picture
($|V(P) \cap V(F)| = 2928$ and $|E(P) \cap E(F)| = 1296$).
The differential curation of the two datasets - the Physical dataset is inferred
predominantly from experimental techniques while the Functional dataset is inferred predominantly from computational techniques -
along with the presence of many missing (true negatives) and spurious (false positives) interactions, 
give rise to these low overlaps.
Though this is an observation from only the two yeast datasets considered here,
it may be worthwhile investigating how far away are we from the ``ideal" picture of physical interactions
being a proper subset of functional interactions in order to make most effective use of the two.

\section{Conclusions}
In this work, we attempt to reconstruct ``sparse" complexes from PPI networks, a problem which has not been explored
in previous works (see the recent survey by~\cite{Li2010}) mainly because of the overused assumption
that complexes form ``dense" regions within the networks. Though this assumption might be valid, relying too much on it
in the wake of insufficient PPI data makes it ineffective to detect sparse complexes. To counter this, we employ
functional interactions, which again has not been tried before.
This approach will particularly be effective in detecting complexes where a significant portion of the physical interactions
are unknown or unreliable, for instance, in human.
In addition to these, we also develop some theory around ``complex derivability" (the CE score) 
that could be useful for developing new computational methods.
For example, in the future we will looking at devising a new computational approach that selectively uses 
functional interactions by treating them differently from physical interactions.

\section*{Acknowledgements}

The authors would like to thank Limsoon Wong (NUS), Hufeng Zhou (NUS) and Gary Bader (UToronto) for
insightful discussions, Guimei Liu (NUS) and Shobhit Jain (UToronto) for providing the scoring softwares,
and the reviewers for their detailed comments and suggestions.


%

\end{document}